\documentclass[a4paper,10pt,twoside]{just}

\usepackage{multicol}
\usepackage{graphicx}
\usepackage{booktabs}
\usepackage{amssymb,bm,mathrsfs,bbm,amscd}
\usepackage[tbtags]{amsmath}
\usepackage{lastpage}
\usepackage{CJK}

\begin{document}
\begin{CJK*}{GBK}{song}

\fancyhead[c]{\small  10th International Workshop on $e^+e^-$ collisions from $\phi$ to $\psi$ (PhiPsi15)}
 \fancyfoot[C]{\small PhiPsi15-\thepage}

\footnotetext[0]{Received 30 Nov. 2015}

\title{The Belle II Experiment and SuperKEKB Upgrade}

\author{%
      Boqun Wang$^{1;1)}$\email{boqunwg@ucmail.uc.edu} (for the Belle II Collaboration)
}
\maketitle

\address{%
$^1$ Department of Physics, University of Cincinnati,  Cincinnati,  OH 45221,  U.S.
}

\begin{abstract}
  The Belle II / SuperKEKB experiment is an $e^+e^-$ collider running
  at the $\Upsilon(4S)$ resonance energy to produce B meson pairs. As
  an upgrade of the Belle / KEKB experiment, it will start physics
  data taking from 2018 and with $\sim 40$ times luminosity, its goal
  is to accumulate 50 $ab^{-1}$ of $e^+e^-$ collision data. Now the
  upgrade of the sub-detector systems is on-going in KEK. The physics
  program has a wide range of areas, including searches for direct
  CPV, Lepton Flavour Violation and dark matter. In this proceedings,
  we review the current upgrade status of Belle II and SuperKEKB and
  introduce some physics opportunities at this facility.

\end{abstract}



\begin{multicols}{2}

\section{Introduction}

The so-called B factory is an $e^+e^-$ collider running at the
$\Upsilon(4S)$ resonance energy to produce B meson pairs. The major B
factories are Belle running at KEKB in Japan and BaBar running at
PEP-II in US. The total data set collected by these two facilities is
$\sim 1.5$ $ab^{-1}$ of $e^+e^-$ collision data. With that data
sample, they've reached physics achievements in areas like the CKM
angle measurement, $|V_{cb}|$ and $|V_{ub}|$ measurement, semileptonic
and leptonic B decays, rare B decays, $\tau$ physics, $D^0$ mixing and
CPV, B physics at the $\Upsilon(5S)$, two-photon physics and new
resonances~\cite{bphys}.

For searching the New Physics (NP), which is physics beyond the
Standard Model (SM), the Belle / KEKB experiment will be upgraded to
Belle II / SuperKEKB~\cite{belleii}. The upgraded detector is planning
to take $\sim 50$ $ab^{-1}$ of $e^+e^-$ collision data. The SuperKEKB
asymmetric electron positron collider can provide a clean environment
for producing B meson pairs via $\Upsilon(4S)$ resonance decay. Its
designed luminosity is $8 \times 10^{35} cm^{-2} s^{-1}$, which is
about 40 times larger than the KEKB collider. The 50 $ab^{-1}$ overall
integrated luminosity corresponds to 55 billion $B \overline{B}$
pairs, 47 billion $\tau^+\tau^-$ pairs, and 65 billion $c
\overline{c}$ states.

In this article, we introduce the Belle II / SuperKEKB experiment, the
current status and future plan of the experiment, and the
opportunities for New Physics.

\section{SuperKEKB}

Many sub-systems of the SuperKEKB accelerator need to be upgraded for
achieving the 40 times luminosity compared with KEKB. The most
important part is the beam size. By using the so-called nano-beam
technology, the beam bunches are significantly squeezed as shown in
Fig.~\ref{fig:nano}. The beam energies of positron and electron will
be changed slightly, from 3.5 GeV / 8 GeV to 4 GeV / 7 GeV, to achieve
a less boosted center-of-mass system.

\begin{center}
\includegraphics[width=\linewidth]{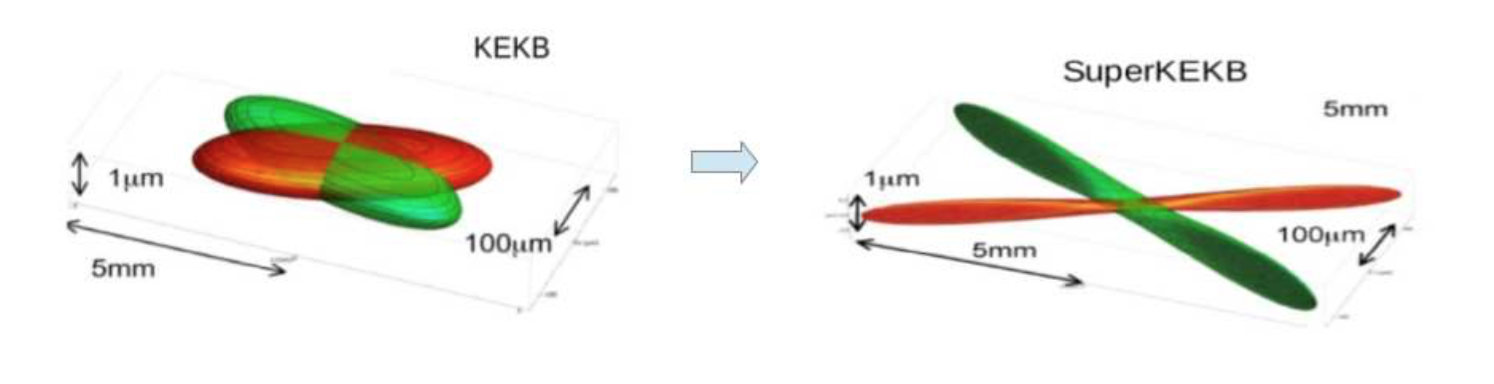}
\figcaption{\label{fig:nano} The beam size comparison between KEKB
  (left) and SuperKEKB (right).}
\end{center}

\section{Belle II detector}

As shown in Fig.~\ref{fig:belle-ii}, most sub-detectors of Belle will
be upgraded for Belle II. This includes the newly designed
vertex detection system (PXD and SVD), a drift chamber with longer
arms and smaller cells, a completely new PID system which consists of
TOP detector at the barrel and ARICH detector in the forward end, the
electro-magnetic calorimeter (ECL) with upgraded crystals and
electronics, and upgraded $K_L$-$\mu$ detection system (KLM). More
details will be introduced in following sections.

\begin{center}
\includegraphics[width=0.8\linewidth]{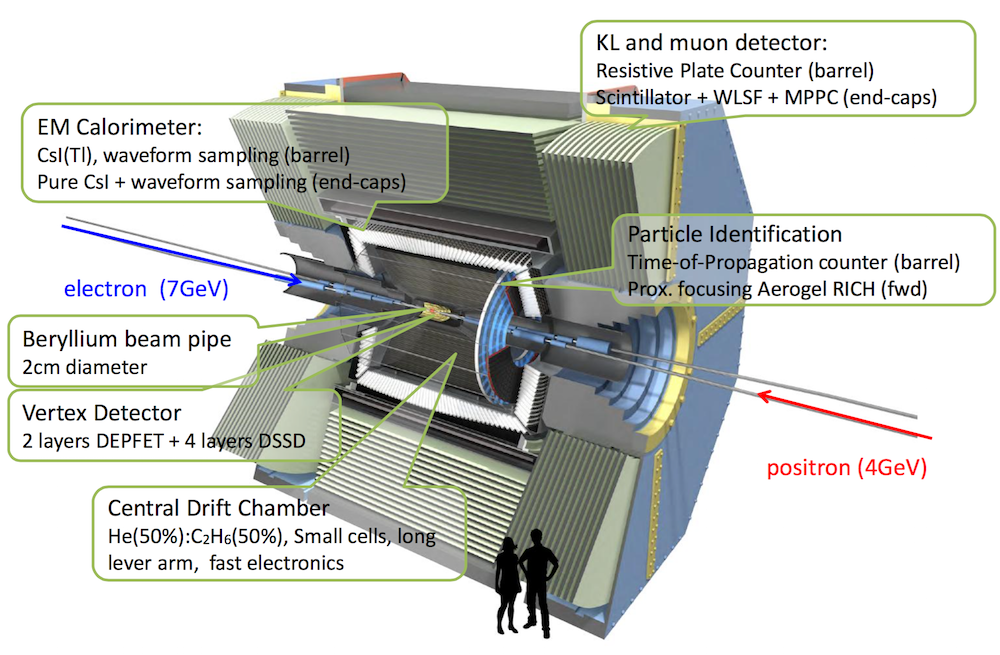}
\figcaption{\label{fig:belle-ii}  Overview of the Belle II detector and
  its sub-detectors.}
\end{center}

\subsection{VXD}

The vertex detector consists of two parts: PXD in the inner part and
SVD in the outer part. PXD consists of two layers of DEPFET (DEPleted
p-channel Field Effect Transistor) and SVD consists of four layers of
DSSD (Double Sided Strip Detectors), as shown in Fig.~\ref{fig:vxd}.
These two sub-detectors combined should have a good vertex resolution
for charged tracks. Now the system integration is on-going, and a beam
test for SVD just finished in the summer of 2015.

\begin{center}
\includegraphics[width=0.48\linewidth]{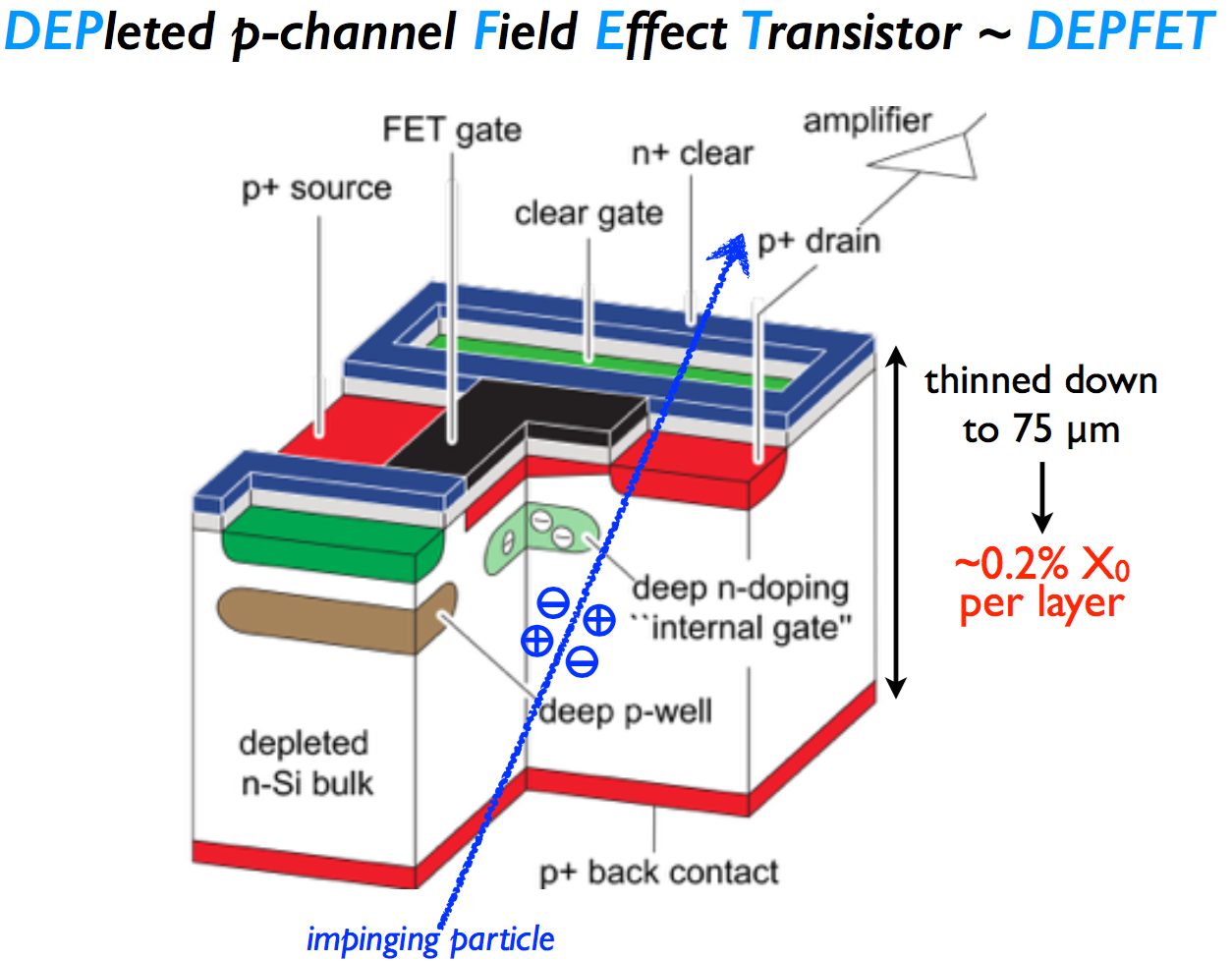}
\includegraphics[width=0.48\linewidth]{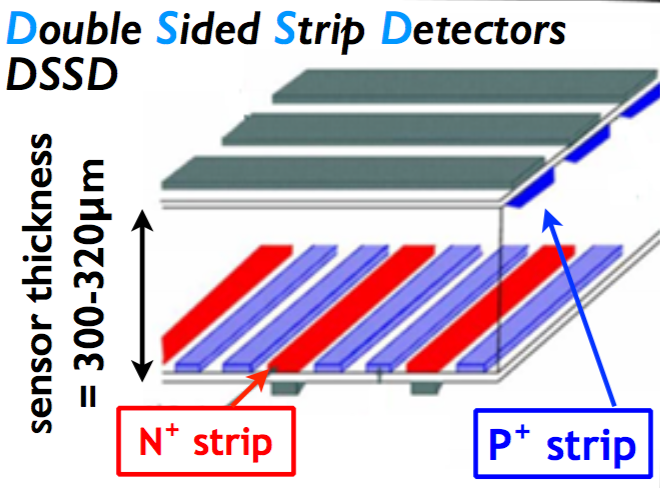}

\figcaption{\label{fig:vxd}  The structure of the DEPFET (left) and
  DSSD (right).}
\end{center}

\subsection{CDC}

As the main tracking device for charged tracks, the CDC in Belle II is
larger than that in Belle and it has a smaller cell size, as shown in
Fig.~\ref{fig:cdc}. This should improve the momentum and dE/dx
resolution. The stringing for the CDC was finished in January of 2014
with 51456 wires and now it's being commissioned with cosmic rays.

\begin{center}
  \includegraphics[width=0.8\linewidth]{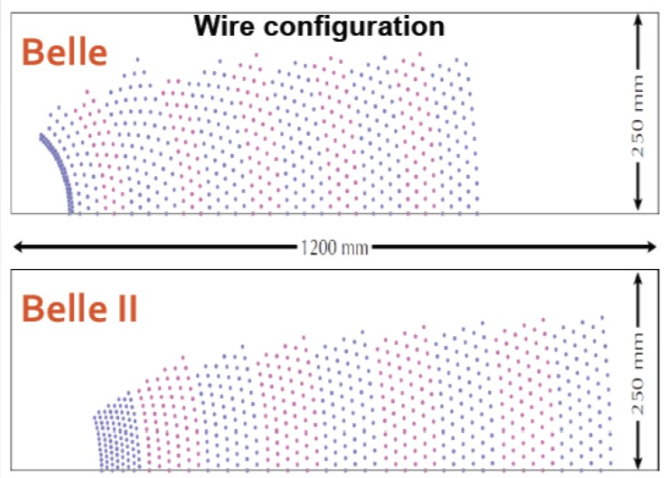}
  \figcaption{\label{fig:cdc} The comparison of CDC wire
    configurations between Belle (top) and Belle II (bottom).}
\end{center}

\subsection{TOP}

The imaging Time of Propagation sub-detector (TOP or iTOP) will be
used for particle identification in the barrel region of Belle II.
There are 16 TOP modules, and each one consists of two quartz bars,
one mirror, one prism, and an array of photo-detectors to collect
Cerenkov photons generated by charged tracks going through the
radiator, as shown in Fig.~\ref{fig:top-schematic}. To distinguish
between kaons and pions, the photo-detectors have excellent position
and timing resolution. This is achieved by using MCP-PMTs and new
waveform sampling electronics.

\begin{center}
  \includegraphics[width=\linewidth]{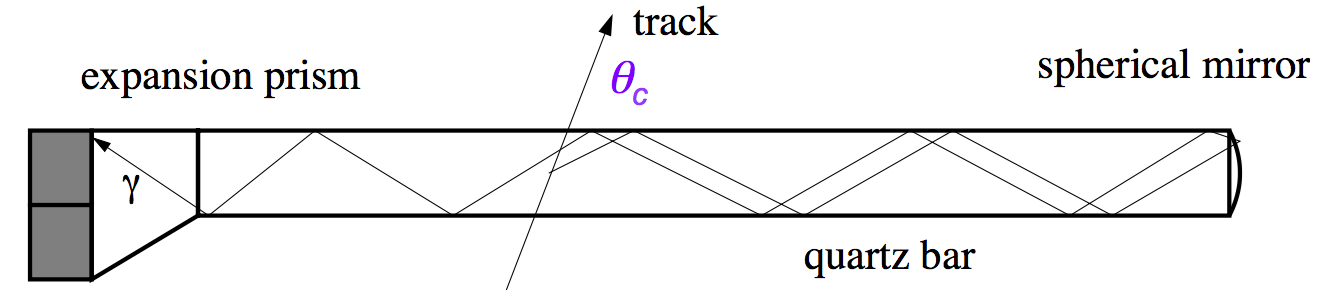}
  \figcaption{\label{fig:top-schematic} The structure of the TOP
    detector.}
\end{center}

TOP modules have been tested during the beam test at SPring-8 at LEPS
in 2013, and good agreement between data and MC simulation has been
obtained, with timing requirement $\sim$ O(100 ps), as shown in
Fig.~\ref{fig:top-beam-test}.

\begin{center}
  \includegraphics[width=0.8\linewidth]{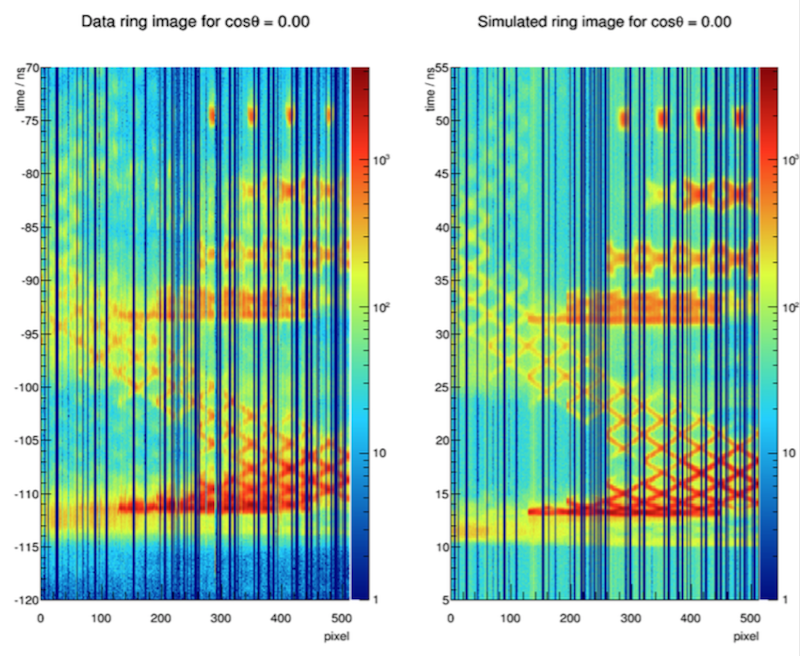}
  \figcaption{\label{fig:top-beam-test} The TOP beam test data (left)
    and simulated MC sample (right).}
\end{center}

The assembly of the TOP modules is on-going in KEK, and all modules
should be finished by spring of 2016. The commissioning with
cosmic ray is under way.

\subsection{ARICH}

Aerogel Ring Imaging Cerenkov (ARICH) detector will be used for
particle identification in the forward end-cap. Two layers of aerogel
with different indices of refraction will be used to improve the
resolution of the detector. For readout, 420 Hybrid Avalanche Photo
Detectors (HAPD), each with 144 channels, will be used, as shown in
Fig.~\ref{fig:arich}.

\begin{center}
  \includegraphics[width=0.3\linewidth]{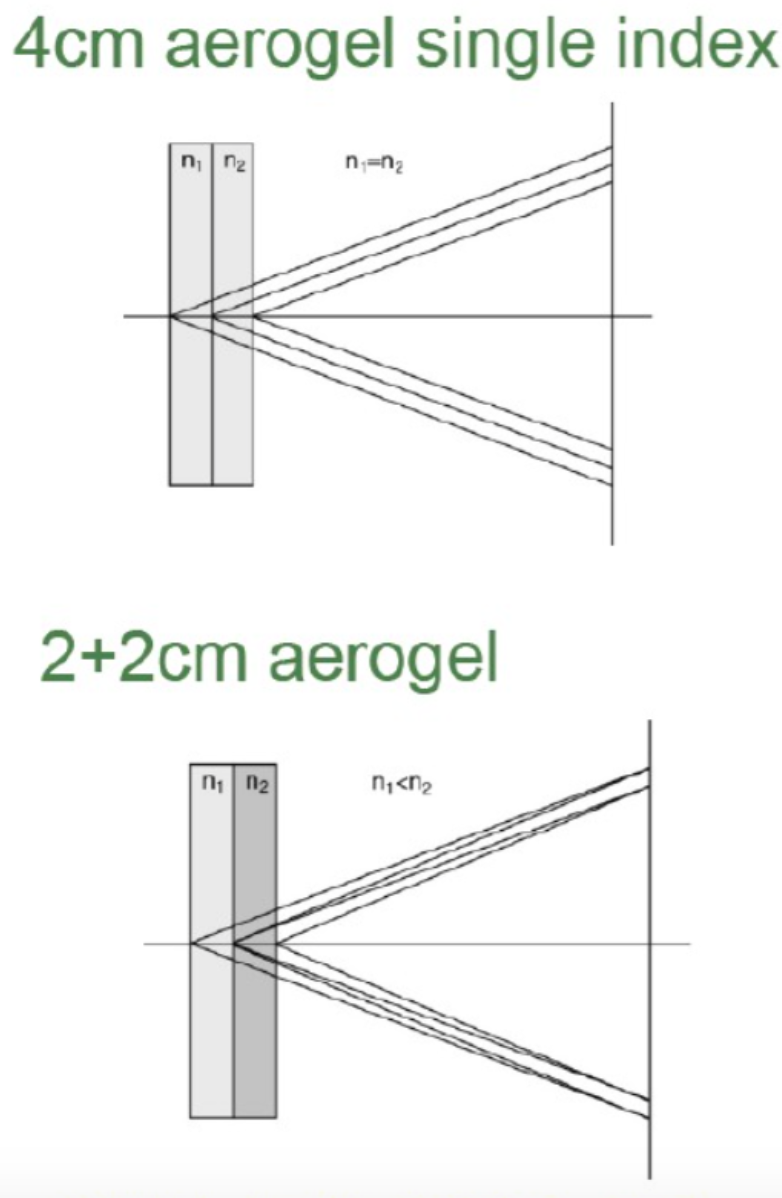}
  \includegraphics[width=0.6\linewidth]{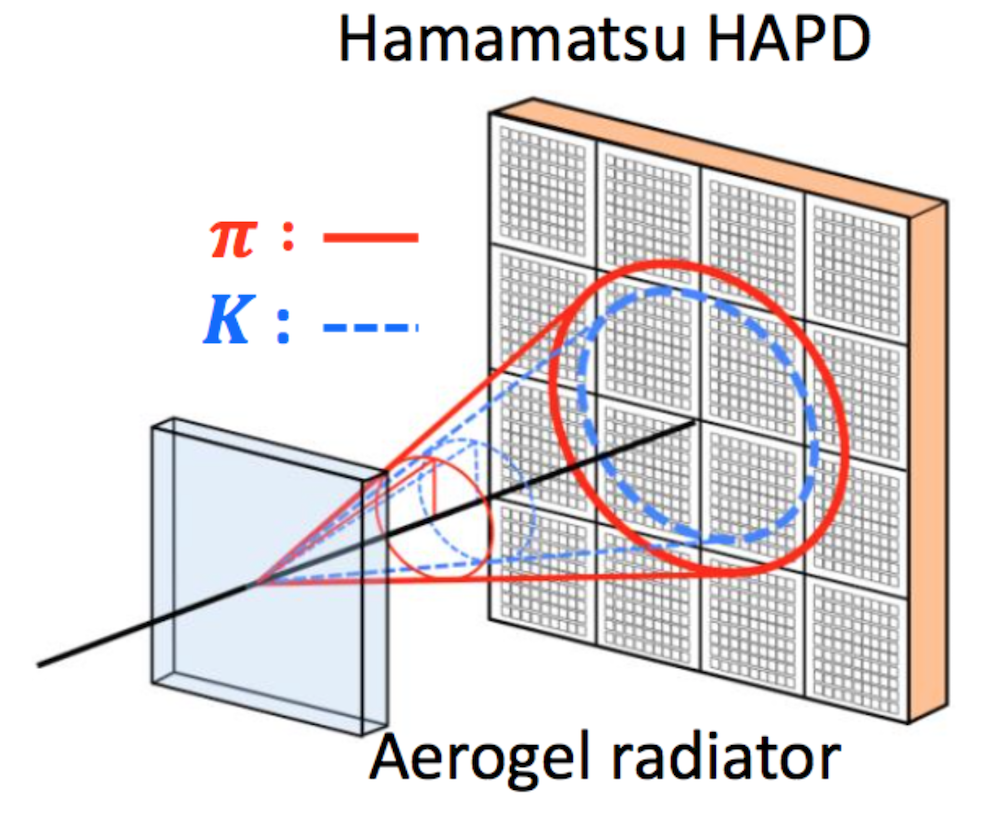}
  \figcaption{\label{fig:arich} The focusing mechanism (left) and the
    structure (right) of ARICH.}
\end{center}

\subsection{ECL}

For the upgrade of the ECL detector, the crystals in barrel side will
be re-used and the crystals in end-cap will be refurbished. New
electronics, such bias filter and waveform sampling will be used for
the upgraded detector. Now the cosmic ray test is underway. The
expected performance for ECL is shown in Fig.~\ref{fig:ecl}.

\begin{center}
  \includegraphics[width=0.8\linewidth]{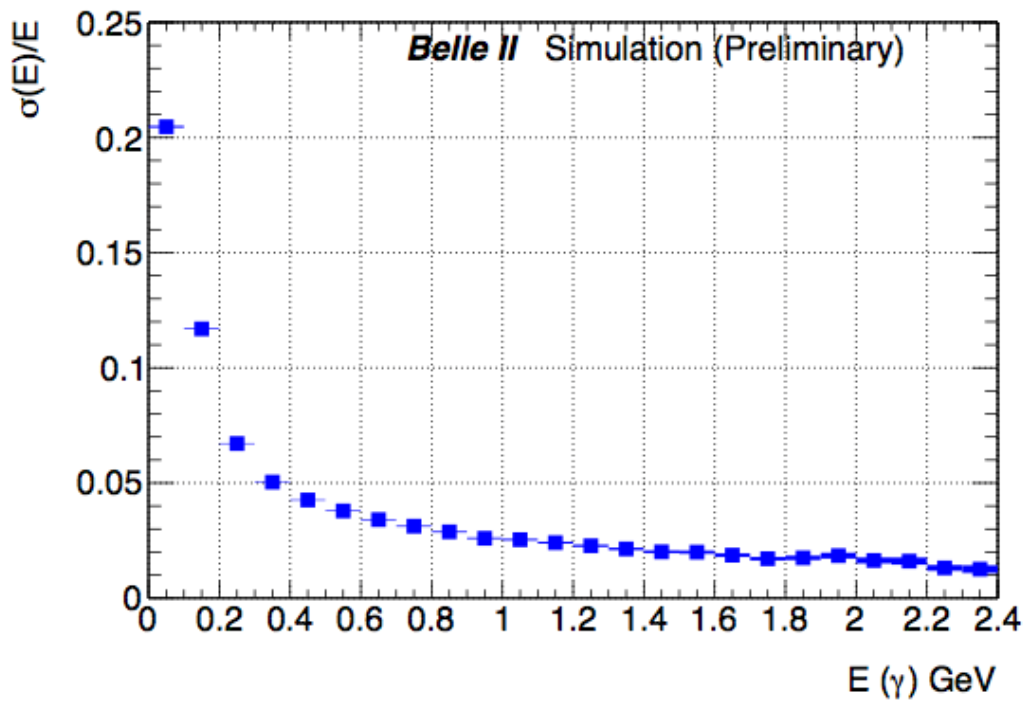}
  \figcaption{\label{fig:ecl} The expected performance of the ECL detector.}
\end{center}

\subsection{KLM}

The endcaps and the inner layers of the barrel RPCs of KLM will be replaced
with scintillators due to increased backgrounds expected in Belle II,
as shown in Fig.~\ref{fig:klm}. The barrel KLM was the first sub-detector
to be installed in Belle II.

\begin{center}
  \includegraphics[width=0.8\linewidth]{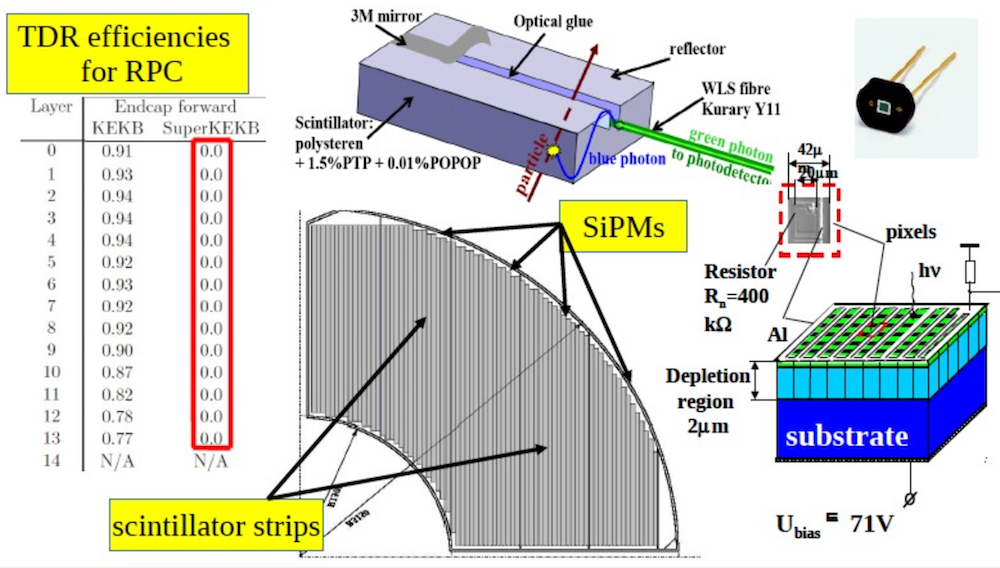}
  \figcaption{\label{fig:klm} The structure of the KLM detector. }
\end{center}

\section{Physics opportunities}

There should be many potential signals for new physics in Belle II,
such as the flavor changing neutral currents, probing charged Higgs,
new sources of CPV, Lepton Flavour Violation decays, and searches for
a dark photon. With the much larger data set compared with Belle and
BaBar, Belle II will contribute to the search of the new physics,
together with the upgraded LHCb. For example, the CKM Unitarity
Triangle should be significantly improved, as shown in
Fig.~\ref{fig:ckm}~\cite{ckm}.

\begin{center}
  \includegraphics[width=0.8\linewidth]{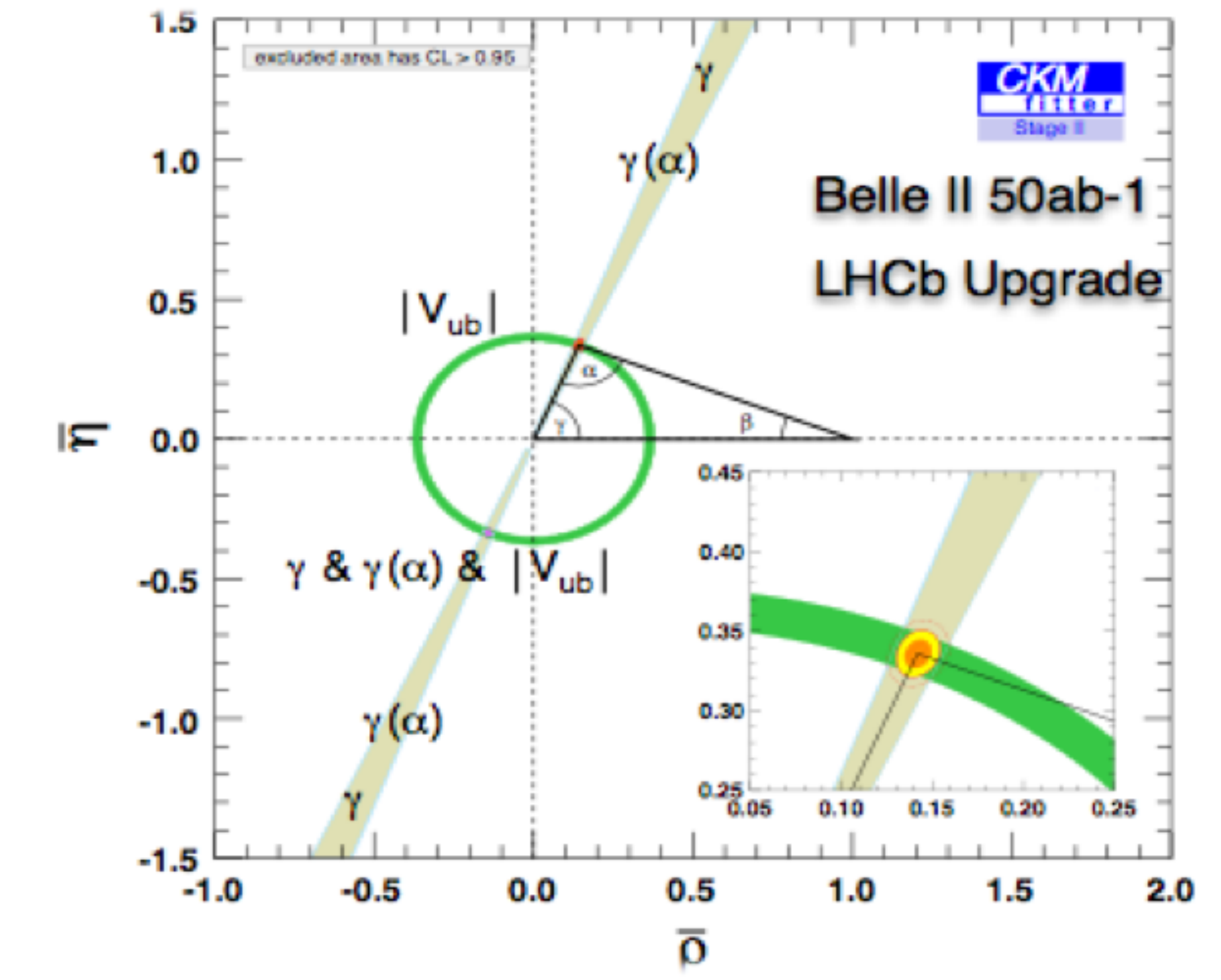}
  \figcaption{\label{fig:ckm} The predicted accuracy of CKM Unitarity
    Triangle with data taken by LHCb and Belle II, from~\cite{ckm}. }
\end{center}

\subsection{Direct CPV in $D^0 \rightarrow \phi \gamma, \rho^0
  \gamma$}

The direct CPV in radiative decays can be enhanced to exceed
1\%~\cite{cpv1}. The $A_{CP}$ for $D^0 \rightarrow \phi \gamma$ could
be up to 2\%, and the $A_{CP}$ for $D^0 \rightarrow \rho^0 \gamma$
could be up to 10\%. The decay for $D^0 \rightarrow \phi \gamma$ was
first observed by Belle with 78 $fb^{-1}$ of data, with the relative
error on yield of about 25\%~\cite{cpv2}. With 50 $ab^{-1}$ of data,
the $A_{CP}$ sensitivity will be reduced to 1\%.

% \begin{center}
%   \includegraphics[width=0.8\linewidth]{direct-cpv.png}
%   \figcaption{\label{fig:cpv} The decay of $D^0 \rightarrow \phi \gamma, \rho^0
%   \gamma$.}
% \end{center}

\subsection{$D^0 \rightarrow \gamma \gamma$}

The branching fraction of the decay $D^0 \rightarrow \gamma \gamma$ is
predicted by SM as $\sim 10^{-8}$. The upper limit by BaBar with 470
$fb^{-1}$ of data is $2.2 \times 10^{-6}$ with 90\% CL~\cite{d0gg}, as
shown in Fig.~\ref{fig:d0gg}.

With 50 $ab^{-1}$ of data by Belle II, the upper limit could be
improved to $\sim 2 \times 10^{-8}$, if it scales with luminosity L,
or $\sim 2 \times 10^{-7}$, if it scales with $\sqrt{L}$.

\begin{center}
  \includegraphics[width=0.8\linewidth]{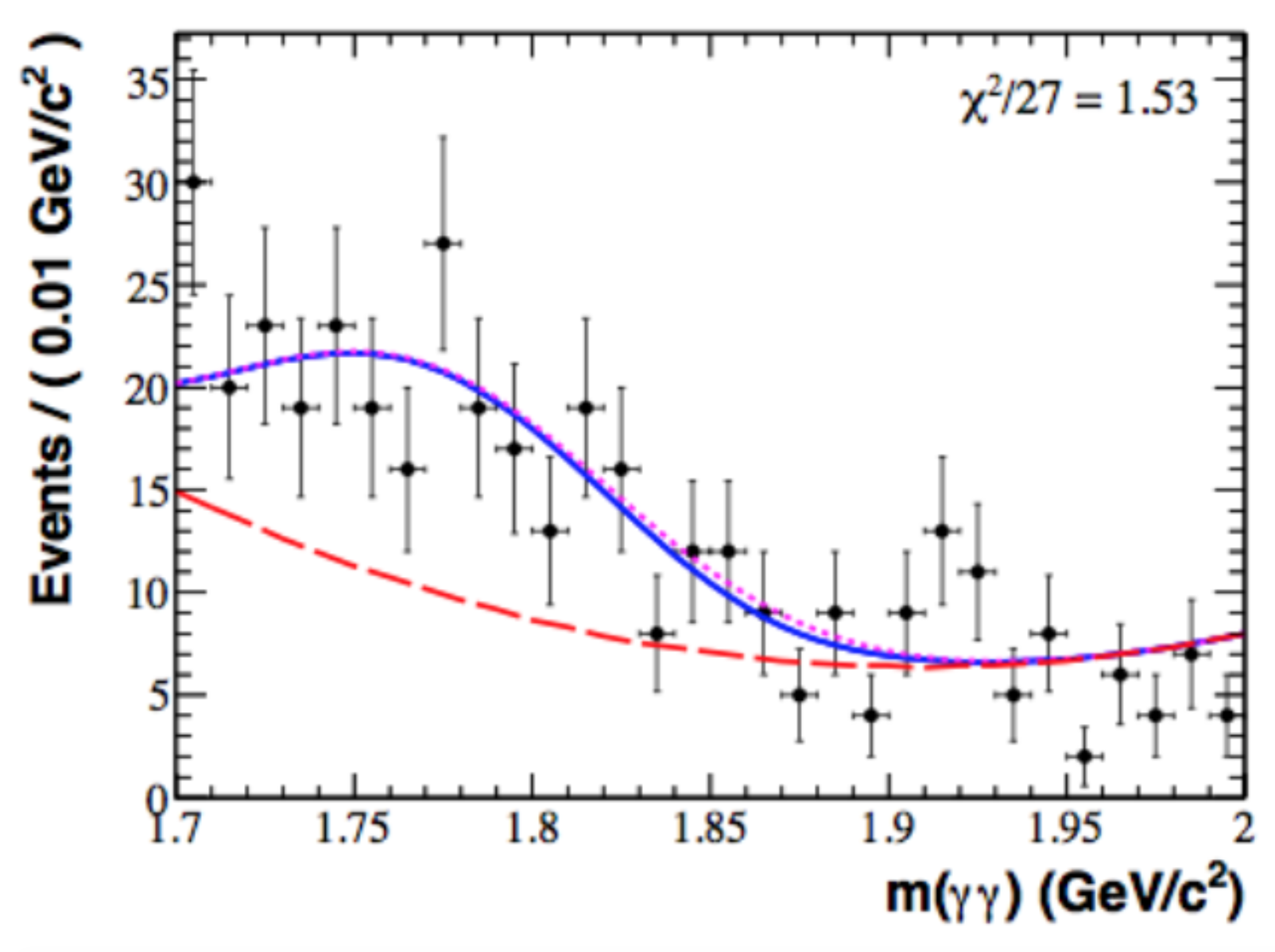}
  \figcaption{\label{fig:d0gg} The decay of $D^0 \rightarrow \gamma
    \gamma$ from BaBar \cite{d0gg}. }
\end{center}

\subsection{$\tau$ Lepton Flavour Violation}

The Lepton Flavour Violation decays are highly suppressed by SM, with
a branching fraction of $\sim 10^{-25}$. But they could be enhanced in
certain New Physics scenarios. Belle has searched for
LFV~\cite{lfv1}~\cite{lfv2}, but no trace of NP has been found. As
shown in Fig.~\ref{fig:lfv}, the red dots show the sensitivity for
some LFV decays in Belle II~\cite{lfv}. The branching fraction of the
decays is within the capability of the Belle II experiment.

\begin{center}
  \includegraphics[width=0.8\linewidth]{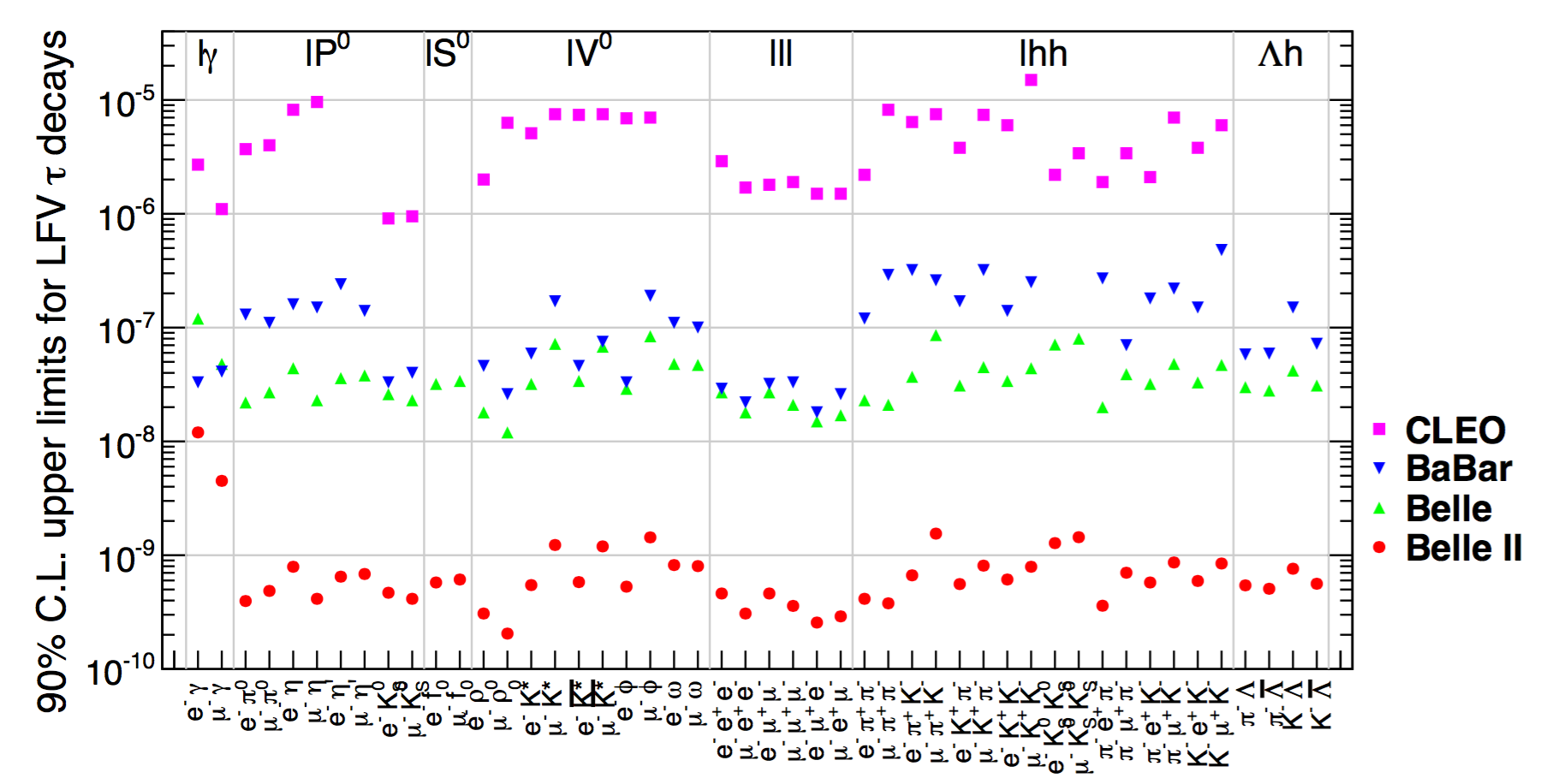}
  \figcaption{\label{fig:lfv} The comparison of LFV upper limit by
    different experiments for different decay channels, from \cite{lfv}}
\end{center}

\subsection{Dark sector}

The dark photon $A'$ is one candidate for dark matter that could be
searched for at an accelerator. Its mass is predicted to be in the
range of MeV to GeV. There are two ways to detect a dark photon:
probing leptonically decaying dark photons through mixing, or probing
sub-GeV dark matter in invisible decays. The upper limits of dark
photon measurement for different experiments are shown in
Fig.~\ref{fig:dark}~\cite{dark}. Belle II has an advantage to search
for dark photon $A'$ with much higher integrated luminosity.

\begin{center}
  \includegraphics[width=0.8\linewidth]{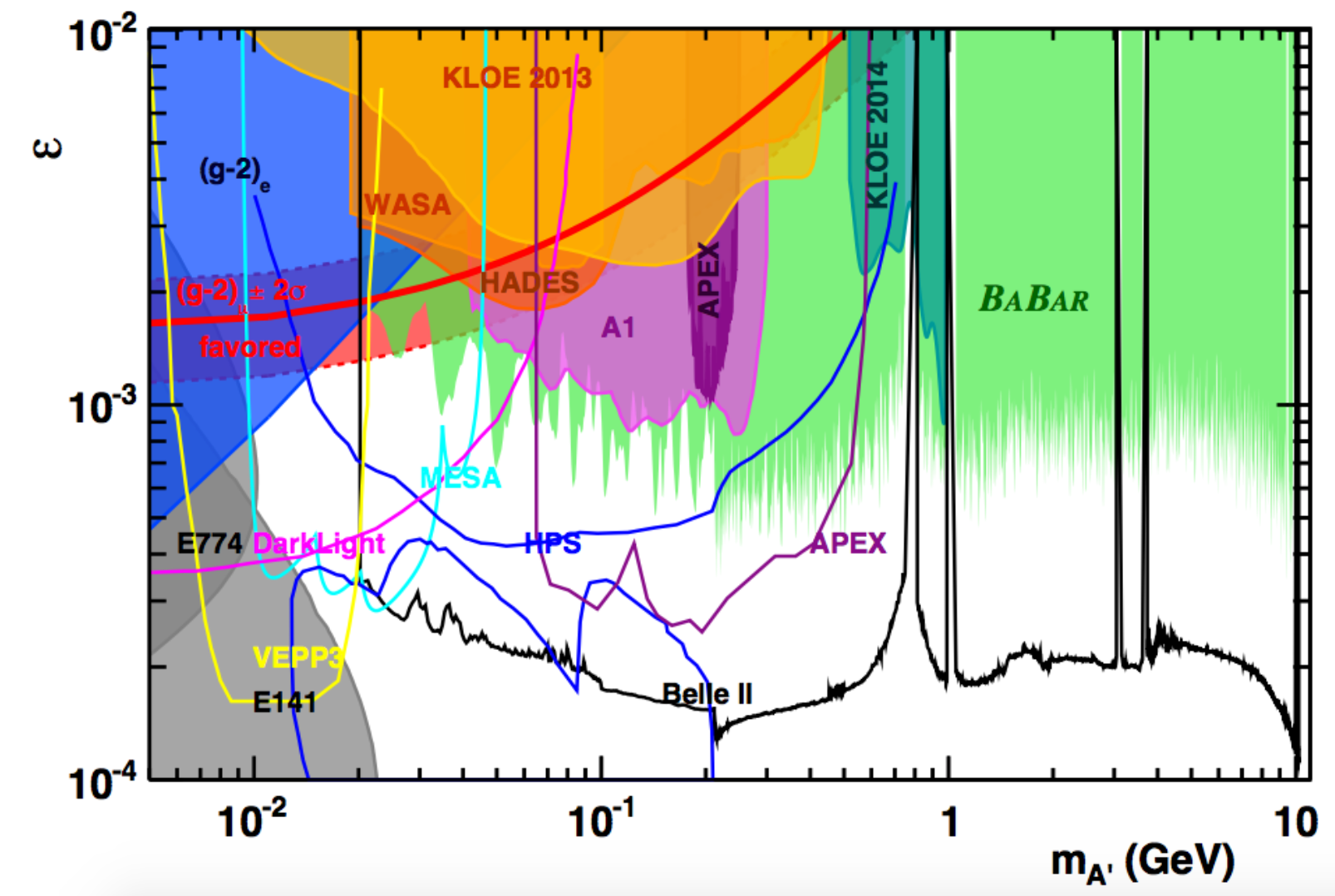}
  \figcaption{\label{fig:dark} The upper limit by different
    experiments for the searching of the dark photon, from \cite{dark}}
\end{center}

\section{Schedule}

The SuperKEKB accelerator is now at the final stage of construction
and the upgrade of the Belle II detector is on-going. As shown in
Fig.~\ref{fig:schedule}, there are three phases in the commission and
operation of Belle II. In phase 1, which begins in early 2016, the
commissioning of various components will start without rolling-in the
detector. In phase 2, which begins in the middle of 2017, Belle II
detector will be partly commissioned to take test physics data without
the vertex detector. Finally, in phase 3, which is expected to start
at the end of 2018, the Belle II detector with full apparatus will
take physics data.

\begin{center}
  \includegraphics[width=0.8\linewidth]{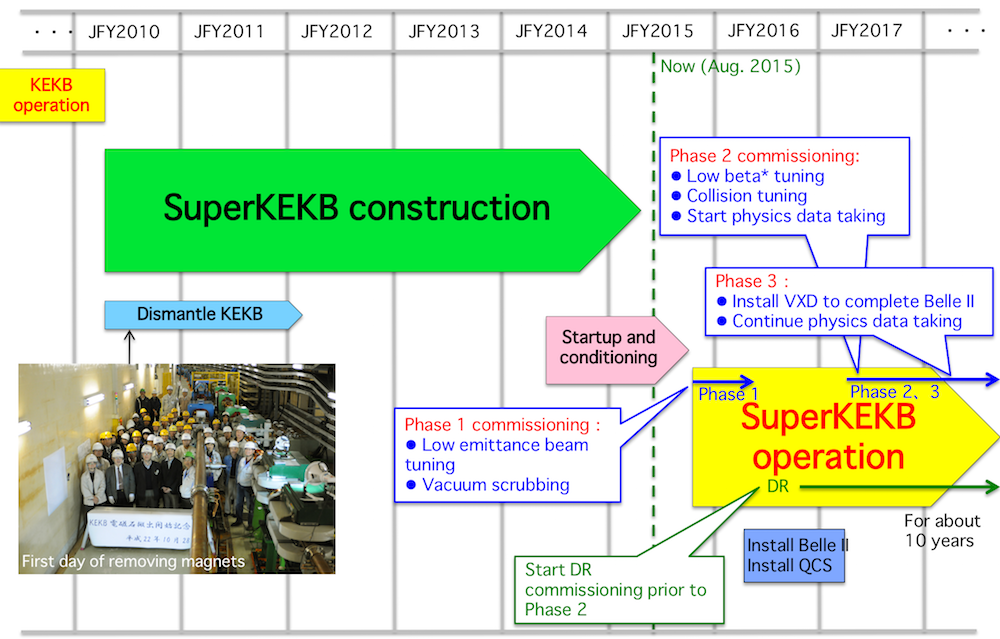}
  \figcaption{\label{fig:schedule} The current Belle II commissioning
    schedule.}
\end{center}

The plan for instantaneous and integrated luminosity is shown in
Fig.~\ref{fig:lum}. According to this plan, the target integrated
luminosity of 50 $ab^{-1}$ will be achieved by 2024.

\begin{center}
  \includegraphics[width=0.8\linewidth]{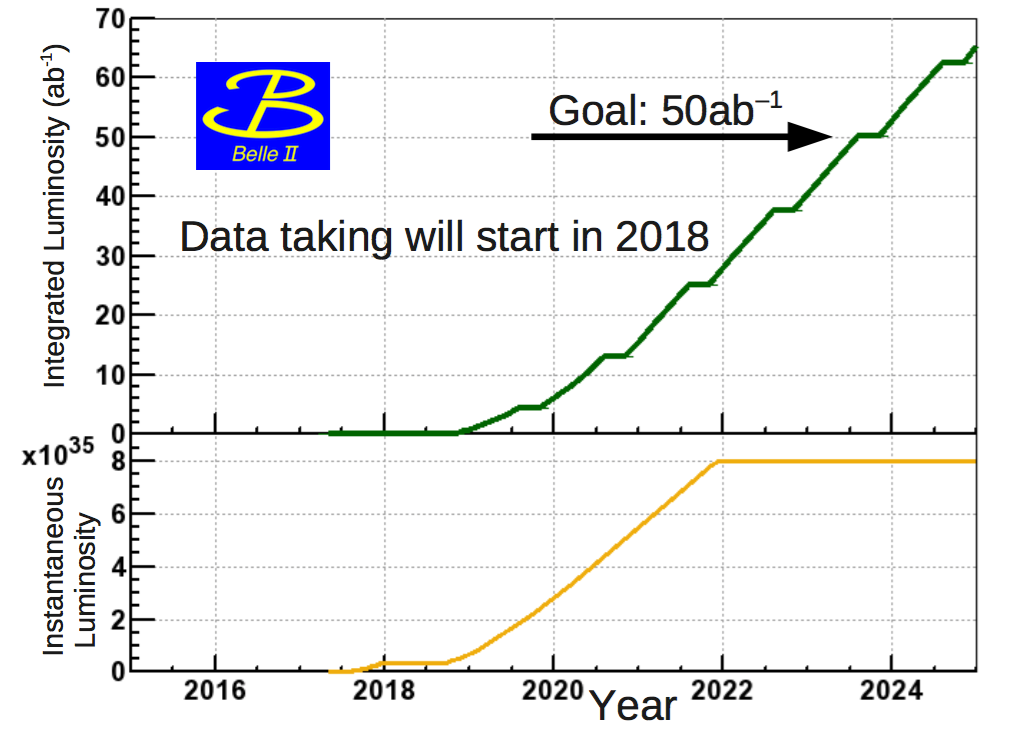}
  \figcaption{\label{fig:lum} The plan for Belle II data taking. }
\end{center}

\section{Summary}

Belle and BaBar as B factories have made many contributions for
flavour physics. As an upgrade, the Belle II / SuperKEKB experiment
could play an important role in the search for New Physics. With the
upgraded accelerator and detector, the experiment will have much
higher luminosity and much better performance for detecting final
state particles.

With the much larger data set collected with the upgraded detector,
Belle II has a rich physics program, which makes it possible to study
the channels with missing energy and neutral particles in the final
states. Now the accelerator and detector are under construction, and
the physics data taking will start at the end of 2018.

\acknowledgments{I'd like to thank the organizers of the PhiPsi15
  conference for inviting me to give this talk. I'd also like to
  express my gratitude to the Belle II collaboration and the HEP group
  of the University of Cincinnati. }

\end{multicols}

\vspace{10mm}

\vspace{-1mm}
\centerline{\rule{80mm}{0.1pt}}
\vspace{2mm}

\begin{multicols}{2}

\end{multicols}

\clearpage

\end{CJK*}
\end{document}